\documentclass[%
reprint,
superscriptaddress,
 amsmath,amssymb,
 aps,
]{revtex4-1}

\usepackage{graphicx}
\usepackage{dcolumn}
\usepackage{bm}
\usepackage{multirow}
\newcommand{\ignore}[1]{}
\usepackage{lineno}

\begin{document}

\title{Experimental Authentication of Quantum Key Distribution with Post-quantum Cryptography}

\author{Liu-Jun Wang}
\thanks{Liu-Jun Wang and Kai-Yi Zhang contribute equally to this work.}
\affiliation{Hefei National Laboratory for Physical Sciences at Microscale and Department of Modern Physics, University of Science and Technology of China, Hefei 230026, China}
\affiliation{Shanghai Branch, CAS Center for Excellence and Synergetic Innovation Center in Quantum Information and Quantum Physics, University of Science and Technology of China, Shanghai 201315, China}
\affiliation{School of Physics and Astronomy and Yunnan Key Laboratory for Quantum Information, Yunnan University, Kunming 650500, China}
\author{Kai-Yi Zhang}
\thanks{Liu-Jun Wang and Kai-Yi Zhang contribute equally to this work.}
\affiliation{Department of Computer Science and Engineering, Shanghai Jiao Tong University, Shanghai 200240, China}
\affiliation{Shanghai Qizhi Institute, Shanghai 200232, China}
\author{Jia-Yong Wang}
\affiliation{CAS Quantum Network Co., Ltd, Shanghai 201315, China}
\author{Jie Cheng}
\affiliation{QuantumCTek Co., Ltd, Hefei 230088, China}
\author{Yong-Hua Yang}
\affiliation{CAS Quantum Network Co., Ltd, Shanghai 201315, China}
\author{Shi-Biao Tang}
\affiliation{QuantumCTek Co., Ltd, Hefei 230088, China}
\author{Di Yan}
\affiliation{Department of Computer Science and Engineering, Shanghai Jiao Tong University, Shanghai 200240, China}
\author{Yan-Lin Tang}
\affiliation{QuantumCTek Co., Ltd, Hefei 230088, China}
\author{Zhen Liu}
\affiliation{Department of Computer Science and Engineering, Shanghai Jiao Tong University, Shanghai 200240, China}
\author{Yu Yu}
\affiliation{Department of Computer Science and Engineering, Shanghai Jiao Tong University, Shanghai 200240, China}
\affiliation{Shanghai Qizhi Institute, Shanghai 200232, China}
\author{Qiang Zhang}
\affiliation{Hefei National Laboratory for Physical Sciences at Microscale and Department of Modern Physics, University of Science and Technology of China, Hefei 230026, China}
\affiliation{Shanghai Branch, CAS Center for Excellence and Synergetic Innovation Center in Quantum Information and Quantum Physics, University of Science and Technology of China, Shanghai 201315, China}
\author{Jian-Wei Pan}
\affiliation{Hefei National Laboratory for Physical Sciences at Microscale and Department of Modern Physics, University of Science and Technology of China, Hefei 230026, China}
\affiliation{Shanghai Branch, CAS Center for Excellence and Synergetic Innovation Center in Quantum Information and Quantum Physics, University of Science and Technology of China, Shanghai 201315, China}


\maketitle

\textbf{Quantum key distribution (QKD) can provide information theoretically secure key exchange even in the era of quantum computer \cite{Bennett84,EKERT91,Scarani09}. However, QKD requires the classical channel to be authenticated, and the current method is pre-sharing symmetric keys \cite{Fung2010}. For a QKD network of $n$ users, this method requires $C_n^2 = n(n-1)/2$ pairs of symmetric keys to realize pairwise interconnection. In contrast, with the help of mature public key infrastructure (PKI) and post-quantum cryptography (PQC) with quantum resistant security, each user only needs to apply for a digital certificate from certificate authority (CA) to achieve efficient and secure authentication for QKD. We only need to assume the short-term security of the PQC algorithm to achieve the long-term security of the distributed keys. Here, we experimentally verified the feasibility, efficiency and stability of the PQC algorithm in QKD authentication, and demonstrated the advantages when new users join the QKD network. Using PQC authentication we only need to believe the CA is safe, rather than all trusted relays. QKD combined with PQC authentication will greatly promote and extend the application prospects of quantum safe communication.}



Recently, Google claimed to have achieved quantum supremacy \cite{arute2019quantum}, a major milestone towards the development of quantum computers. Quantum computing can efficiently solve classical hard problems such as integer factorization and discrete logarithms and demonstrates its quadratic speedup (over classical algorithms) in solving unstructured search problems \cite{shor1994algorithms,grover1996fast}, which poses a serious threat to the security of classical cryptographic algorithms based on the complexity of these problems. Boudot et al.~\cite{Boudot20} recently announced the factoring of RSA-240, a RSA number of  240 decimal digits or 795 bits, as well as solved a discrete logarithm of the same size. New records of this type are constantly being refreshed as the performance of computer hardware increases over time. In the era of quantum computing, there are two kinds of reliable information security mechanisms: one is quantum cryptography \cite{Gisin02}, which mainly includes quantum key distribution; the other is post-quantum cryptography, such as lattice-based cryptography and code-based cryptography, which cannot be effectively cracked by the currently known quantum computing algorithms.

Quantum key distribution is unconditionally secure based on the principle of quantum mechanics. With realistic devices, the security of QKD can also be guaranteed \cite{xu2020secure}. The experiments and practical applications of QKD have drastically developed. The secure key rate reaches 26.2 Mbps at a channel loss of 4 dB (equivalent to a 20-km-long optical fiber) \cite{islam2017provably}, and the maximum key distribution distance through practical optical fiber has exceeded 500 km \cite{PhysRevLett.124.070501,fang2020implementation}. Micius satellite has realized entanglement-based repeaterless quantum key distribution between two places on the ground at a distance of 1120 km \cite{yin2020entanglement}. Through trusted relay, several quantum communication networks have been built \cite{Peev09,Chen10,Sasaki11,Froehlich13,wang2014field,PhysRevLett.120.030501}, and the ``Beijing-Shanghai backbone'' quantum communication network spans 2200 km.

Nowadays, the hardness of most public key cryptography are based on integer factorization and discrete logarithm problems that are difficult or intractable for conventional computers. However, Shor's \cite{shor1994algorithms} quantum algorithm can achieve an exponential speedup in solving these mathematical problems. In 2016, NIST published a report on Post-quantum Cryptography \cite{CJL+2016} anticipating that a quantum computer is likely to be built by 2030 that breaks 2000-bit RSA in a few hours, and therefore renders the current public-key infrastructure insecure. As a result, in the same year NIST initiated the ``Post-Quantum Cryptography Standardization'' process by announcing a call for proposals of quantum resistant cryptographic primitives including public key encryption, digital signature and key exchange algorithms. And the process is expected to release the standardization documents by 2024. 

\begin{figure*}
\includegraphics{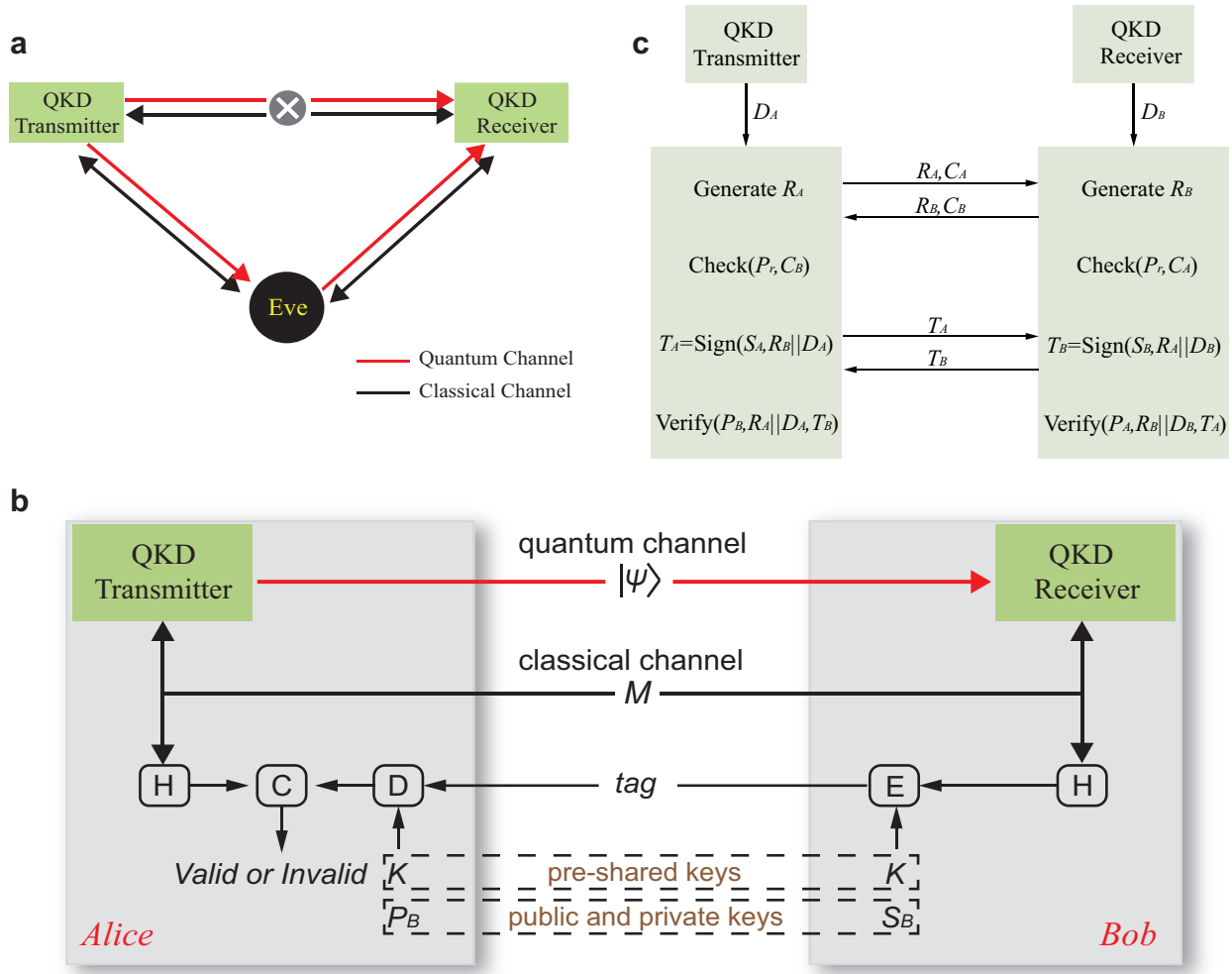}
\caption{\label{fig:MITM_Flow_PQC} \textbf{Schematic of man-in-the-middle attack and flow diagram of post-quantum cryptography authentication.} \textbf{a,}  As a middleman, Eve pretends to be a legitimate party. He cuts off the quantum channel, reconnects the legitimate parties respectively, and carries out the intercept-resend attack. \textbf{b,}  $|\psi\rangle$: quantum signals; $M$: classical messages; H: hash function, we used SM3 hash algorithm in the experiment; E: encryption (signature); D: decryption; C: comparison; $K$: pre-shared symmetric keys, $P_B$: Bob's public key, $S_B$: Bob's private key. The figure shows that Alice authenticates Bob's identity. In the experiment, we implemented two-way authentication, that is, Bob also authenticates Alice's identity. \textbf{c,} $D$: digest; $R$: random nonce, we generate them by Intel chips; $C$: valid certification; $P_r$ public key of certificate authority; $||$ : concatenate two bit strings ; $T$: tag (or signature) of concatenation of $R$ and $D$; $S$: private key; $P$: public key.}
\end{figure*}

Quantum key distribution includes the quantum channel that transmit photons and the classical channel used in post data processing. The unconditional security of QKD does not require the classical channel to be confidential, but requires it to be authenticated, otherwise there will be a man-in-the-middle attack. Combined with the intercept-resend attack, the attacker can completely obtain the keys of both parties without being discovered, as shown in Fig.~\ref{fig:MITM_Flow_PQC}a.

The processes of QKD that requires authentication includes: basis sifting, error correction verification, random number transfer needed for privacy amplification, and final key verification \cite{Fung2010}. QKD requires two-way authentication between the two parties.

The current secure authentication method is to pre-share a small amount of symmetric seed keys and encrypt (sign) and decrypt (verify) the hash value of classical messages, as shown in Fig.~\ref{fig:MITM_Flow_PQC}b. Later, the generated quantum key can be used for authentication. This way can guarantee the information theoretical security, however when the number of QKD network users is large, this method is not easy to operate and has the following problems. On the one hand, for a network with arbitrary two users connected, if the number of users is $n$, then the number of pre-shared key pairs $m$ is
\begin{eqnarray}
m = C_n^2 = \frac{n(n-1)}{2}%
\label{eq:one}.
\end{eqnarray}
Symmetric keys are generally pre-shared by face-to-face. When the number of users is relatively large, the burden of pre-sharing keys is heavy and inefficient. For example, if $n$=100, then $m$=4950. At the same time, each user needs to store the authentication key pairs with all other users. The storage, synchronization and management of so many key pairs will increase the complexity and security risk of the network. One solution is to use a trusted relay to form a star-type network, each user only connects and pre-shares one key pair with the trusted relay \cite{Froehlich13,hughes2013networkcentric}, but this reduces the interconnection between users. Moreover, when new users join a QKD network, they need to pre-share symmetric keys with the trusted relay or the original users on demand. If the new user's QKD task is urgent, it may be too late to distribute the authentication key pairs.

Another type of secure authentication method is using the post-quantum public key algorithm and PKI \cite{10.1007/978-3-642-38616-9_9}, as shown in Fig.~\ref{fig:MITM_Flow_PQC}b, c. Each user gets a digital certificate signed by a trusted certification center, which contains his/her identity, public key and other items required by the PKI standard. For a network of $n$ users, the number of digital certificates issued is $n$. If a new user joins the QKD network, he/she only needs to obtain a digital certificate. Therefore, the authentication based on the public key algorithm can solve the problems of pre-sharing symmetric keys. As long as the PQC algorithm is secure during the authentication process, the security of QKD will not be affected, even if the PQC is cracked after authentication, so we only need to assume the short-term security of PQC. This is different from using PQC algorithm for confidentiality or key distribution, which will require long-term security of the PQC algorithm. Here, we verifies the application of PQC in QKD authentication, which greatly improves the operability and efficiency of QKD authentication process.

\begin{figure}
\includegraphics{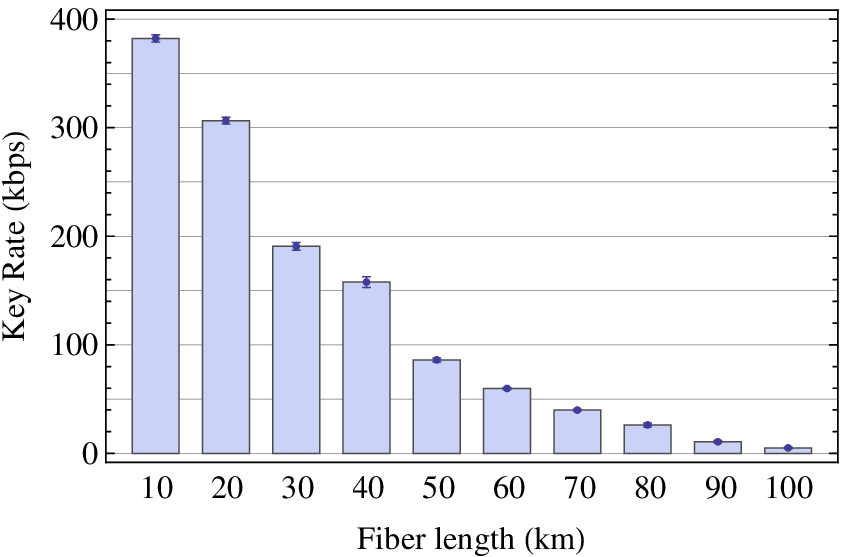}
\caption{\label{fig:KR_Length} 
\textbf{The secure key rate as a function of fiber length when QKD is authenticated by the PQC algorithm.} They are the average values in five minutes. The error bar represents a standard deviation.}
\end{figure}

We realized the application of PQC in the QKD point-to-point link, with fiber distances from 10 km to 100 km. Figure~\ref{fig:KR_Length} shows the key rates as a function of fiber length. It can be seen that the key rates decrease exponentially with fiber length, which is in consistent with the theoretical expectation. We compared the key rates at the same fiber length using the pre-shared key authentication and the post-quantum algorithm authentication, and the two were consistent within the statistical error. This is because the execution time of post-quantum algorithm authentication is less than 1 ms (see Methods), far less than one authentication cycle of the QKD system, which is 1s. In the experiment, we also deliberately set PQC to feedback that the authentication failed, and as a result, the QKD system will discard the keys for these periods.

\begin{figure*}
\includegraphics{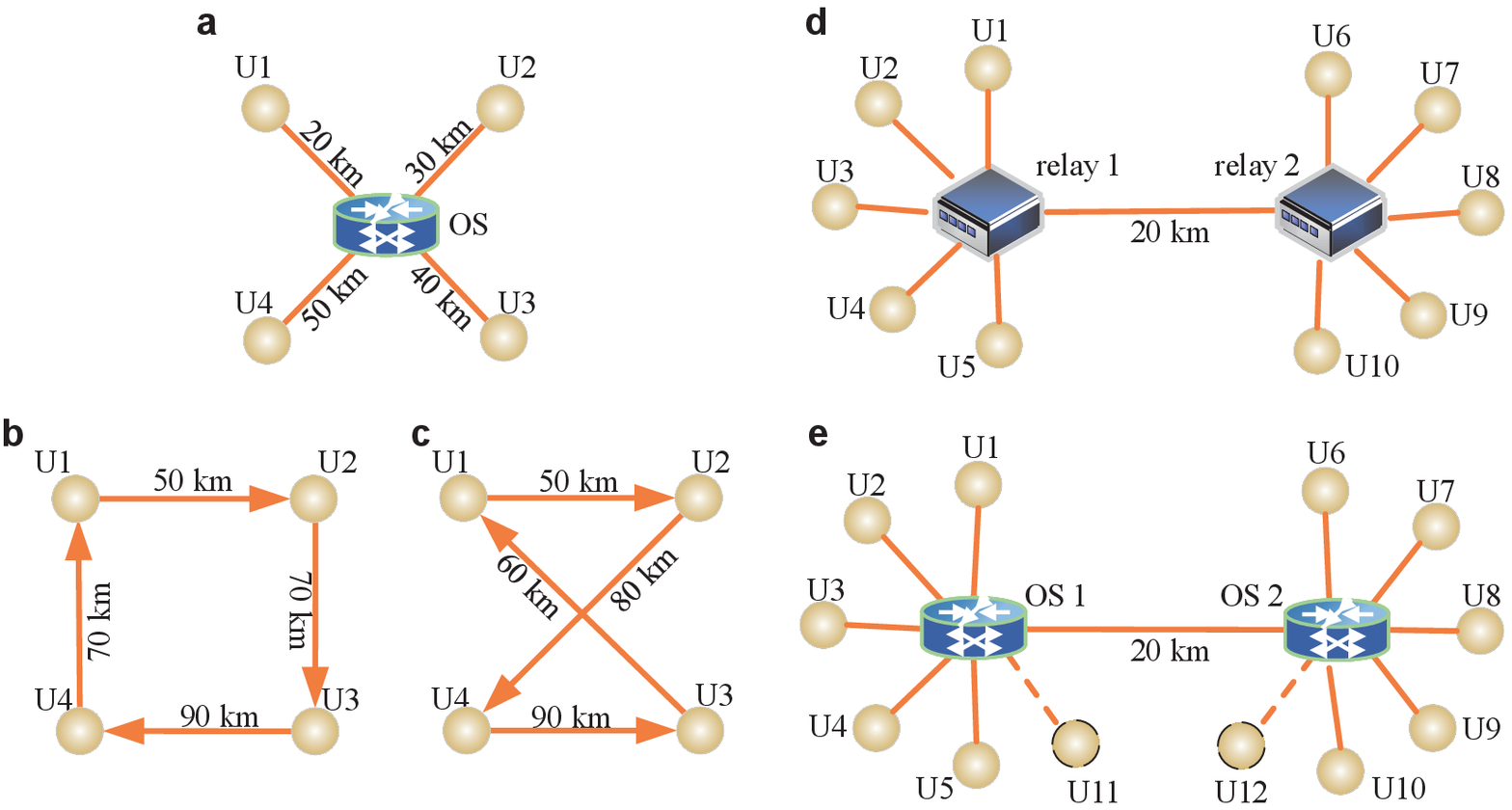}
\caption{\label{fig:networks} \textbf{PQC authentication in QKD networks.}
\textbf{a,} All-pass QKD network. Four users are connected to each other through an optical switch. \textbf{b,} Ring network. \textbf{c,} Cross network. The actual distance between any two users is the sum of their respective distances from the optical switch. \textbf{d,} A 10-node QKD metropolitan area network composed of two relay networks. \textbf{e,} Trusted relays are replaced with optical switches to form an all-pass network. U11 and U12 are new users. See the text for the distance from each user to the trusted relay (optical switch).}
\end{figure*}

QKD networks can be generally divided into two types: all-pass network and trusted relay network. For the all-pass network, users are connected by optical switches (OS). In order to achieve arbitrary connection between users, each user must have a QKD transmitter and a receiver. We built an all-pass network for four users, connected by an optical switch, as shown in Fig.~\ref{fig:networks}a. It can realize two typical topological relationships, one is ring connection and the other is cross connection, as shown in Fig.~\ref{fig:networks}b and Fig.~\ref{fig:networks}c respectively. We verified the application of PQC authentication in these two kinds of all-pass networks. The experimental results are shown in Table~\ref{tab:all-pass}. We note that because the performance of different QKD devices are not exactly the same, their key rates and QBERs will be different under the same fiber lengths. Using PQC authentication, we also demonstrated the QKD relay network (see Supplementary Fig. 1 and Tabel I).

\begin{table}
\caption{\label{tab:all-pass}Key rates and QBERs of the QKD all-pass network authenticated by PQC algorithm.}
\begin{ruledtabular}
\begin{tabular}{cccdc}
Connection&Length (km)&Loss (dB)&\multicolumn{1}{c}{\textrm{Key rate (kbps)}}&QBER\\
\hline
\multicolumn{5}{c}{\textrm{(a) Ring network}}\\
\hline
U1-U2 & 50 & 11.26 & 72.16 & 0.751\%\\
U2-U3 & 70 & 15.35 & 20.17 & 1.140\%\\
U3-U4 & 90 & 18.81 & 10.52 & 0.883\%\\
U4-U1 & 70 & 15.4 & 30.58 & 0.647\%\\
\hline
\hline
\multicolumn{5}{c}{\textrm{(b) Cross network}}\\
\hline
U1-U2 & 50 & 11.21 & 68.65 & 0.779\%\\
U2-U4 & 80 & 16.31 & 19.45 & 1.014\%\\
U4-U3 & 90 & 18.46 & 9.71 & 0.786\%\\
U3-U1 & 60 & 12.15 & 76.82 & 0.517\%\\
\end{tabular}
\end{ruledtabular}
\end{table}

The above results verify the feasibility of PQC algorithm for QKD network authentication. In order to demonstrate the efficiency of PQC authentication, we built two trusted relay networks and connected them to simulate the QKD metropolitan area network. They can be located on both sides of a city. Each relay network contains 5 user nodes, and a total of 10 users in the entire network, as shown in Fig.~\ref{fig:networks}d.

When using pre-shared key authentication, the trusted relay is usually needed to manage pre-shared keys at the cost of reducing the interconnection. With PQC authentication, the trusted relay can be replaced with an optical switch to realize arbitrary interconnection. Each user only needs one digital certificate for authentication, instead of pre-sharing $C_{10} ^ 2 = 45$ pairs of symmetric keys, as shown in Fig.~\ref{fig:networks}e. The interconnectivity of the QKD network has been greatly improved. To illustrate this point, in the experiment, we compared the QKD results of three pairs of users U1-U3, U5-U6, and U8-U10 in two cases, as shown in Table~\ref{tab:two_relay_OS}. Moreover, PQC authentication only needs to assume that the certificate authority is safe, reducing the security dependence on multiple trusted relays, which can improve the actual security of the entire network.

\begin{table*}
\caption{\label{tab:two_relay_OS}Comparison of key rates and QBERs between relay network and all-pass network. R1 and R2 stand for realy 1 and realy 2 in Fig.~\ref{fig:networks}(e), respectively. The fiber length between two users in the all-pass network is the sum of the fiber lengths of the links between the two users in the relay network.}
\begin{ruledtabular}
\begin{tabular}{ccccdc}
\multicolumn{2}{c}{\textrm{Connection}}&Length (km)&Loss (dB)&\multicolumn{1}{c}{\textrm{Key rate (kbps)}}&QBER\\ 
\hline
\multicolumn{6}{c}{\textrm{(a) Relay network}}\\
\hline
\multirow{2}{*}{U1-U3} &U1-R1&10&2.69&363.59&0.648\% \\
&R1-U3&30&6.70&194.32&0.761\% \\ 
\hline
\multirow{3}{*}{U5-U6} &U5-R1&20&3.99&293.53&0.752\% \\
&R1-R2&20&4.08&288.16&0.475\% \\ 
&R2-U6&20&4.11&288.74&0.364\% \\ 
\hline
\multirow{2}{*}{U8-U10} &U8-R2&10&2.62&287.47&0.511\% \\
&R2-U10&10&2.66&333.06&0.529\% \\ 
\hline
\hline
\multicolumn{6}{c}{\textrm{(b) All-pass network}}\\
\hline
\multicolumn{2}{c}{U1-U3}&40&9.02&90.83&0.630\% \\
\hline
\multicolumn{2}{c}{U5-U6}&60&12.12&48.00&0.978\% \\
\hline
\multicolumn{2}{c}{U8-U10}&20&5.23&200.87&0.514\% \\
\end{tabular}
\end{ruledtabular}
\end{table*}

In the experiment, two new users U11 and U12 join the QKD network, as shown in Fig.~\ref{fig:networks}e. If pre-shared key authentication is used, for the relay network, new users need to pre-share keys with the relay, and can only perform QKD with the relay, but not with other users. For the all-pass network, each new user needs to pre-share 10 pairs of symmetric keys with 10 original users, and 1 pair of keys between the two new users. A total of 21 pairs of keys need to be pre-shared to achieve the connection between any two users. In contrast, if PQC authentication is adopted, trusted relays can be replaced with optical switches. Each new user only needs to apply for one digital certificate, and a total of two digital certificates can realize the connection of any two users. This greatly improves the convenience for new users to access the network and interconnection. After U11 and U12 got digital certificates, we demonstrated the QKD between U11-U2, U11-U7, U12-U4, U12-U9, and U11-U12. The results are shown in Table~\ref{tab:new-users}.

\begin{table}
\caption{\label{tab:new-users}QKD Key rates and QBERs between new users U11 and U12 and original users in the network, and between U11 and U12.}
\begin{ruledtabular}
\begin{tabular}{cccdc}
Connection&Length (km)&Loss (dB)&\multicolumn{1}{c}{\textrm{Key rate (kbps)}}&QBER\\
\hline
U11-U2 & 40 & 8.11 & 139.79 & 0.846\%\\
U11-U7 & 50 & 11.26 & 90.18 & 0.573\%\\
U12-U4 & 40 & 8.11 & 113.42 & 0.792\%\\
U12-U9 & 40 & 8.16 & 101.78 & 0.873\%\\
U11-U12 & 50 & 11.07 & 83.05 & 0.858\%\\
\end{tabular}
\end{ruledtabular}
\end{table}

Finally, we tested the stability of PQC authentication with a pair of QKD devices. The fiber length is 40 km, and it has been running continuously for 30 hours. The PQC program keeps running normally, and QKD systems continuously generate keys (see Supplementary information). 

Summarizing, We used the lattice-based post-quantum digital signature algorithm Aigis-Sig, combined with PKI, to achieve efficient and quantum secure authentication of QKD. Since the Aigis-Sig algorithm is highly computationally efficient, it does not affect the performance of QKD, such as the key rate. We experimentally verified the feasibility of its application in metropolitan QKD relay network and all-pass network. With PQC authentication, the trusted relay in the QKD network can be replaced with an optical switch. Each user only needs to apply for a digital certificate through PKI to realize the direct connection between any two users. When a new user joins the network, he/she only need to obtain a digital certificate, instead of distributing symmetric keys with all other users, and they can immediately establish a QKD connection. Compared with the pre-shared key authentication, PQC authentication has obvious operability and efficiency advantages. Moreover, if the number of trusted relays is less, the security dependence on trusted relays in the network can be reduced, thus improving the security of the entire QKD network. We have also verified the long-term stability of PQC authentication.

\subsection*{\label{Methods}Methods}
In the experiment, we used the BB84 protocol combined with decoy state method \cite{Ma05}, with polarization encoding. The system operating frequency was 625 MHz, and single photon detectors based on InGaAs avalanche photodiodes were used. The QKD transmitter and the QKD receiver were synchronized by periodic pulsed light. The synchronous light transmitted with the quantum signal light via a single optical fiber through wavelength-division multiplexing. The QKD systems used SM3 hash algorithm to generate digest values of 256 bits for the messages to be authenticated, and output them to PQC program. The finite-key effect is considered in the data processing.

The PQC algorithm we used is Aigis-Sig \cite{zhang2020tweaking}, an efficient lattice-based digital signature scheme from variants of Learning With Errors (LWE) \cite{Regev05} and Small Integer Solutions (SIS) \cite{Ajtai96} problems. It has been shown that these two problems are at least as hard as some worst-case lattice problems (e.g., Gap-SIVP) for certain parameter choices\cite{Regev09,Peikert09,GPV08}. Therefore, the post-quantum security of Aigis-Sig algorithm is based on the conjectured quantum resistance of the underlying lattice problems. Furthermore, it has not been found that quantum algorithms have substantial advantages (beyond polynomial speedup) over classical ones in solving lattice problems. 

Our authentication protocol adopts a PKI enhanced with post-quantum secure Aigis-Sig as shown in Fig.~\ref{fig:MITM_Flow_PQC}c. The protocol consists of two phases. In the first phase, the transmitter and the receiver first exchange their own certificates issued by the certificate authority (CA) to each other. Then they use the public key of CA to verify the other public key belongs to its identity. In the second phase, the transmitter and the receiver first use our Aigis-Sig to sign the message digest under their own private keys, then they use the confirmed public keys of the other to verify the correctness of the receiving signatures. Because only the legitimate party has the corresponding private key, it can be confirmed that the message is signed legally. 

In order to prevent the replay attack, we introduce the nonce in our authentication protocol, the nonce is a random number generated by Intel chips. We exchange the nonce in the first phase and concatenate them with the message digest together as our signing message in the second phase. Note that we implemented two-way authentication in QKD data processing. 

We implement PQC algorithm in Win10 64bit, Intel(R) Core(TM) i7-9750H CPU @2.60GHz, 8G RAM. The average CPU cycle of Signature Generation is 459903. The average CPU cycle of Signature Verification is 104337. The signature size is 2445 bytes. The real execution time is less than 1ms.

\begin{acknowledgments}
This work was supported by the National Key R\&D Program of China (Grants No. 2017YFA0304000), the National Natural Science Foundation of China, the Chinese Academy of Sciences (CAS), Shanghai Municipal Science and Technology Major Project (Grant No.2019SHZDZX01), the Anhui Initiative in Quantum Information Technologies, and Yunnan Fundamental Research Project (Grant No.K264202005920) and the Major Science and Technology Project (Grant No.2018ZI002).
\end{acknowledgments}


\bibliography{QKD_PQC}
\bibliographystyle{naturemag}

\subsection*{Supplementary information}
\begin{figure}
\includegraphics{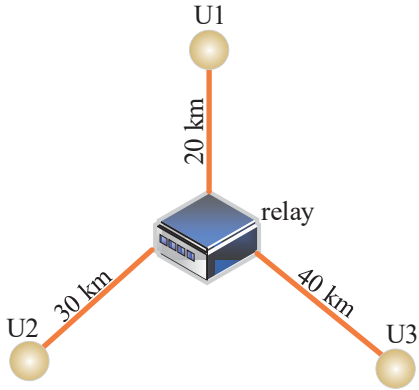}
\caption{\label{fig:relay} 
The QKD network based on trusted relay, U1-U3 represents three user nodes, all of them hold QKD transmitters, relay node places a QKD receiver. The users generate key pairs with each other through the relay.}
\end{figure}

In the relay network experiment, the network includes a relay node and three user nodes, as shown in Fig.~\ref{fig:relay}. Due to the high cost of single photon detectors, relay nodes generally deploy QKD receivers, and user nodes deploy QKD transmitters. The fiber lengths between the relay and the users are typical distances within metropolitan area. Using PQC authentication, the QKD between the relay node and the three users was successfully achieved. The key rates and QBERs are shown in the Table~\ref{tab:relay}, they are the average values in five minutes.

\begin{table}
\caption{\label{tab:relay}Key rates and quantum bit error rates (QBERs) of the QKD relay network authenticated by PQC algorithm.}
\begin{ruledtabular}
\begin{tabular}{cccdc}
Connection&Length (km)&Loss (dB)&\multicolumn{1}{c}{\textrm{Key rate (kbps)}}&QBER\\
\hline
relay-U1 & 20 & 4.03 & 309.55 & 0.704\%\\
relay-U2 & 30 & 6.87 & 226.62 & 0.381\%\\
relay-U3 & 40 & 8.23 & 109.71 & 0.742\%\\
\end{tabular}
\end{ruledtabular}
\end{table}

\begin{figure}
\includegraphics{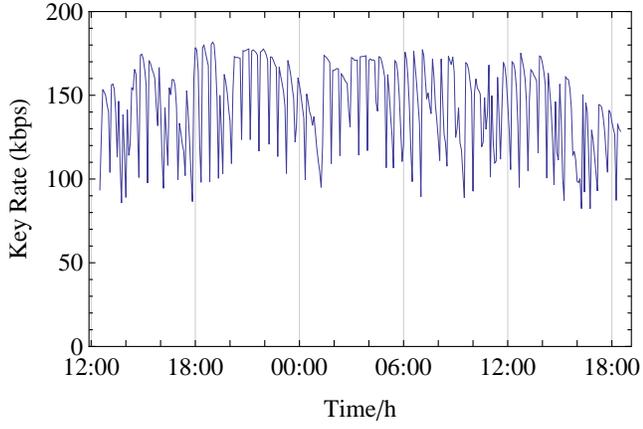}
\caption{\label{fig:stability_keyrate} 
A curve of 30-hour QKD key rates. The fiber distance is 40 km, and each value is an average of 5 minutes. The near periodic fluctuation of the key rate is due to the fact that the QKD system is set to start polarization feedback when the QBER reaches 3\% or the QKD continuous running time reaches 30 minutes, so the data points containing polarization feedback in the key rate statistical period will be lower.}
\end{figure}

\begin{figure}
\includegraphics{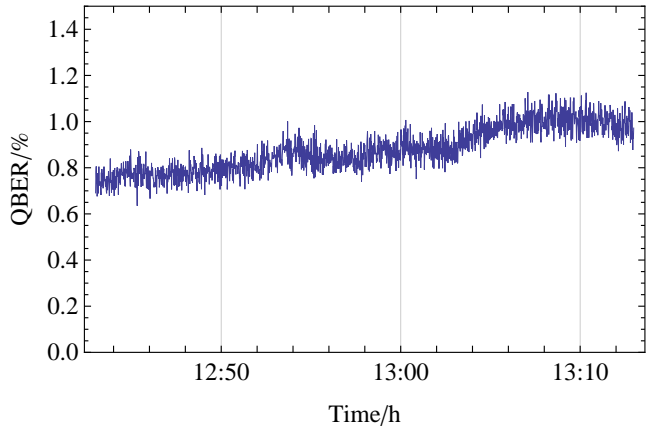}
\caption{\label{fig:stability_QBER} 
A curve of QBER for 30 minutes. One data is recorded per second in the experiment. The average QBER is 0.876\%.}
\end{figure}

In the experiment, we tested the stability of PQC authentication with a pair of QKD devices. The fiber length is 40 km, and it has been running continuously for 30 hours. The PQC program keeps running normally, and QKD systems continuously generate keys, as shown in Fig.~\ref{fig:stability_keyrate}. The secure key rate is in the range of 100 - 180 kbps, and the fluctuations are caused by the QBER reaching 3\% or the continuous running time reaching 30 minutes, which triggers the polarization feedback set by the QKD system. The 30-hour average key rate is 144.1 kbps. It can also be seen from the figure that polarization feedback is more frequent during the daytime (08:00-18:00) than at night (18:00-08:00), because human activities and temperature fluctuations during the day interfere more seriously with the optical fiber. Figure~\ref{fig:stability_QBER} shows a curve of the QBER within a 30-minute operating cycle of a QKD system. One QBER data is recorded every second. The QBER is distributed between 0.65\% and 1.1\%, and the 30-minute average is 0.876\%. The above results show the stability and reliability of PQC authentication applied to QKD.

\end{document}